\documentclass[12pt]{article}
    \newcommand{\be}[1]{\begin{equation}\label{#1}}
    \newcommand{\ba}[1]{\begin{eqnarray}\label{#1}}

    \newcommand{\pa}[1]{\left(#1\right)}
    \newcommand{\paq}[1]{\left[#1\right]}
    
    \newcommand{\av}[1]{\langle#1\rangle}
    \newcommand{\M}{{\rm M_{\rm P}}}
    \def\ee{\end{equation}}
    \def\ea{\end{eqnarray}}

\usepackage{graphicx,latexsym}
\usepackage{graphicx,epsf}
\usepackage{color}
\usepackage{epsfig}
\usepackage[T1]{fontenc}
\usepackage[utf8]{inputenc}
\usepackage{authblk}
\begin{document}
\title{Inflation and Quantum Gravity in a Born-Oppenheimer Context}
\author[1,2]{Alexander Y. Kamenshchik\thanks{Alexander.Kamenshchik@bo.infn.it}}
\author[1]{Alessandro Tronconi\thanks{Alessandro.Tronconi@bo.infn.it}}
\author[1]{Giovanni Venturi\thanks{Giovanni.Venturi@bo.infn.it}}
\affil[1]{Dipartimento di Fisica e Astronomia and INFN, Via Irnerio 46,40126 Bologna,
Italy}
\affil[2]{L.D. Landau Institute for Theoretical Physics of the Russian
Academy of Sciences, Kosygin str. 2, 119334 Moscow, Russia}
\date{}
\renewcommand\Authands{ and }
\maketitle

\begin{abstract}
A general equation, describing the lowest order corrections coming from quantum gravitational effects to the spectrum of cosmological scalar fluctuations is obtained. These corrections are explicitly estimated for the case of a de Sitter evolution.
\end{abstract}


\section{Introduction}
The effects of quantum gravity are supposed to be very small since they are suppressed by the huge value of the Planck mass. They can become essential in the presence of a strong gravitational field or in the very early universe undergoing an inflationary expansion (see e.g. \cite{inflation}). In this letter we would like to study the possible influence of quantum gravitational effects on the spectrum of cosmological fluctuations produced during inflation. Such fluctuations are imprinted in the cosmic microwave background radiation which is one of the main sources of information on the very early universe (see e.g. \cite{cmb}).

The Born-Oppenheimer (BO) approach \cite {BO} has been extensively applied to composite systems such as molecules, which involve two mass, or time, scales. Such an approach has also been suggested for the matter-gravity quantum system where one expects matter to follow semiclassical gravity adiabatically (in the quantum mechanical sense) \cite{BO-cosm,BO-unit,modes}. The plausibility of such an approach relies on the fact that the mass scale of gravity is the Planck mass which is much greater than that of normal matter. 

The above is also related to the semiclassical emergence of time in the matter-gravity quantum system, indeed this has been studied in a mini-superspace model during inflation \cite{BO-cosm}. Conditions were found for the usual time evolution of quantum matter (Schwinger-Tomonaga or Schr\"odinger) to emerge, essentially these are that non-adiabatic transitions (fluctuations) be negligible or that the universe be sufficiently far from the Planck scale.

The scope of this note is to illustrate quantitatively the possible effects of the non adiabatic transitions on the power spectrum associated with cosmological fluctuations. These should not be confused with loop effects obtained by considering self interactions due to the potential of a matter scalar field or higher order (beyond the second) perturbations of the metric and matter field. Let us begin with the following reduced action:
\ba{act}
S&=&\int d\eta\left\{-\frac{\M^2}{2}a'^2+\frac{a^2}{2}\paq{\phi_0'^2-V(\phi_{0})a^2}\right.\nonumber\\
&+&\left.\sum_{i=1,2}\sum_{k\neq 0}^\infty\paq{v_{i,k}'(\eta)^2+\pa{-k^2+\frac{z''}{z}}v_{i,k}(\eta)^2}
\right\}
\ea
where $\M$ is the Planck mass and we consider a homogeneous metric $ds^2=a(\eta)^2\pa{-d\eta^2+\delta_{ij}dx^idx^j}$. This action has been obtained from the usual Einstein action with a minimally coupled scalar field $\phi(\vec x,\eta)$ and a potential $V(\phi)$. The Einstein action is evaluated for a general metric, including the scalar metric perturbations and in the uniform curvature gauge. The scalar field is expanded as $\phi(\vec x,\eta)=\phi_0(\eta)+\delta \phi(\vec x,\eta)$ and terms up to the second order in the perturbations are kept in the total action \cite{MukMald}. The first order contributions can be eliminated by using the equations of motion for the homogeneous parts. Variation with respect to the two remaining metric perturbations and $\delta\phi$ lead to three equations, two of which can be used to eliminate the former in terms of the scalar field perturbation. One then has the Friedmann equations with the back reaction part coming from the scalar field and an equation of motion just for $\delta \phi$. These equations can be obtained directly from the reduced action (\ref{act}) where, instead of $\delta\phi$, we have introduced $v\equiv a\delta\phi$ (with $v_k$ its Fourier mode, henceforth we shall omit the index $i$), which is the gauge-invariant Mukhanov variable \cite{Mukhanov}  in the uniform curvature gauge and $z\equiv \phi_0'/H$ where $'\equiv d/d\eta$ with $H$ the Hubble parameter.  Let us note that $z''/z$ can be expressed in terms of the scale factor and its time derivatives.

The quantization of our system will lead to fluctuations about solutions of the classical equations of motion. We now observe that in obtaining the reduced action (\ref{act}) we have at most kept terms to quadratic order in the field and metric perturbations ($v_{k}$). Therefore, since quantum fluctuations around $z''/z\equiv -m^{2}(\eta)$ occur already multiplied by small field perturbations we shall just retain for it its classical homogeneous value.\\
In the next section we write down the Wheeler-De Witt (WDW) equation, obtain the expression for the vacuum quantum state of the gauge-invariant scalar fluctuations and derive the general equation to describe the quantum gravitational lowest-order corrections to the spectrum of the cosmological fluctuations. In Sec. III this equation is applied to the case of the de Sitter expansion and we make some concluding remarks.  
\section{Wheeler-De Witt equation and the effect of Quantum Gravity on the spectrum of fluctuations} 
The canonical quantization of the Einstein equation associated with the action (\ref{act}) leads to the following WDW equation \cite{DeWitt}
\begin{eqnarray*}
&&\left\{\frac{1}{2\M^2}\frac{\partial^2}{\partial a^2}-\frac{1}{2a^2}\frac{\partial^2}{\partial \phi_0^2}+Va^4\right.\nonumber\\
&&\left.+\sum_{k\neq 0}^{\infty}\paq{-\frac{1}{2}\frac{\partial^2}{\partial v_k^2}+\frac{\omega_k^2}{2}v_k^2}\right\}\Psi\pa{a,\phi_0,\{v_k\}}=0
\end{eqnarray*}
where $\omega_{k}^{2}\equiv k^{2}+m^{2}(\eta)$. We can now perform a BO decomposition of the WDW wave function $\Psi$ into a gravitational part $\psi(a)$ and a matter part $\chi\pa{a,\phi_0,\{v_k\}}$ 
\ba{WDW}
&&\paq{\frac{1}{2\M^2}\frac{\partial^2}{\partial a^2}+\hat H_0^{(M)}+\sum_k \hat H_k^{(M)}}\psi(a)\chi\pa{a,\phi_0,\{v_k\}}\nonumber\\
&&\equiv\paq{\frac{1}{2\M^2}\frac{\partial^2}{\partial a^2}+\hat H^{(M)}}\psi(a)\chi\pa{a,\phi_0,\{v_k\}}=0.
\ea 
Let us observe that our canonical variables are $a$, $\phi_{0}$ and $v_{k}$'s and the resulting wave function of the universe is $\Psi\pa{a,\phi_0,\{v_k\}}=\langle a,\phi_0,\{v_k\}|\Psi\rangle$. On introducing a suitable basis $\chi_{\lambda}\pa{a,\phi_0,\{v_k\}}$ for the matter field one may always write $\Psi\pa{a,\phi_0,\{v_k\}}=\sum_{\lambda}\chi_{\lambda}\pa{a,\phi_0,\{v_k\}}\psi_{\lambda}(a)$. This can further be written as  
\be{dec}
\Psi\pa{a,\phi_0,\{v_k\}}=\psi(a)\sum_{\lambda}\chi_{\lambda}\pa{a,\phi_0,\{v_k\}}c_{\lambda}(a)\equiv \psi(a)\chi\pa{a,\phi_0,\{v_k\}},
\ee
where $\psi(a)$ will satisfy the gravitational WDW equation (see below) involving the averaged matter Hamiltonian \cite{timeTVV}.\\
The matter part can be further decomposed as $\chi=\chi_0\pa{a,\phi_0}\prod_{k\neq 0}^\infty \chi_k\pa{\eta,v_k}\equiv \prod_{k=0}^{\infty}\chi_k$. One then follows the same procedure as previously employed \cite{BO-cosm,BO-unit,modes} obtaining a first set of coupled equations: the gravitational WDW equation is
\be{graveq}
\paq{\frac{1}{2\M^2}\frac{\partial^2}{\partial a^2}+\av{\hat H^{(M)}}}\tilde \psi=
-\frac{1}{2\M^2}\langle\frac{\partial^2}{\partial a^2}\rangle\tilde \psi
\ee
and the matter equation is
\ba{mateq}
&&\tilde\psi^*\tilde\psi\paq{\hat H^{(M)}-\av{\hat H^{(M)}}}\tilde \chi+\frac{1}{\M^2}\pa{\tilde\psi^*\frac{\partial}{\partial a}\tilde \psi}\frac{\partial}{\partial a}\tilde \chi\nonumber\\
&&=\frac{1}{2\M^2}\tilde\psi^*\tilde\psi\paq{\av{\frac{\partial^2}{\partial a^2}}-\frac{\partial^2}{\partial a^2}}\tilde \chi
\ea
where
\be{gder}
\psi=e^{-i\int^{a} \mathcal{A} da'}\tilde\psi,\;  \chi=e^{i\int^{a} \mathcal{A} da'}\tilde\chi, \; \mathcal{A}=-i\langle\chi|\frac{\partial}{\partial a}|\chi\rangle
\ee
with $v_0=\phi_0$, $\av{\hat O}=\langle\tilde \chi|\hat O|\tilde\chi\rangle$ and each mode is individually normalized $\langle \chi_{k}|\chi_{k}\rangle=\int d v_k\chi_k^*\chi_k=1$.\\
If one now multiplies both sides of Eq. (\ref{mateq}) by $\bar \chi_i\equiv\prod_{j\neq i}\tilde \chi_j$ and contracts over all $v_j$ ($j\neq i$) one finally projects out a single Fourier component
\ba{mateqk}
&&\tilde\psi^*\tilde\psi\paq{\hat H_k^{(M)}-\langle \tilde \chi_k|\hat H_{k}^{(M)}|\tilde \chi_k\rangle}\tilde \chi_k+\frac{1}{\M^2}\pa{\tilde\psi^*\frac{\partial \tilde \psi}{\partial a}}\nonumber\\
&&\times \frac{\partial \tilde \chi_k}{\partial a}=\frac{1}{2\M^2}\tilde\psi^*\tilde\psi\paq{\langle \tilde \chi_k|\frac{\partial^2}{\partial a^2}|\tilde \chi_k\rangle-\frac{\partial^2}{\partial a^2}}\tilde \chi_k.
\ea
We may now perform the semiclassical limit for the gravitational wave function $\psi(a)$ by setting 
\be{semicl}
\tilde\psi(a)\sim(\M^{2}a')^{1/2}\exp\pa{-i\int^{a}\M^{2}a' da}
\ee
and obtaining, for Eq. (\ref{graveq}),
\be{geqsc}
-\frac{\M^2}{2}a'^2+\sum_k\av{\hat H_k^{(M)}}=0
\ee
to the leading order. Eq. (\ref{geqsc}) is the Friedmann equation. In such a way the BO decomposition of the wave function of the universe is uniquely determined. Now, on defining $|\chi_{k}\rangle_{s}\equiv e^{-i\int^\eta\langle \tilde \chi_k|\hat H_{k}^{(M)}|\tilde \chi_k\rangle d\eta'}|\tilde \chi_{k}\rangle$, Eq. (\ref{mateqk}) becomes
\ba{evoeq}
\!\!\!\!\!\!\!&&\!\!\!\!\!\!\!i\partial_{\eta}|\chi_{k}\rangle_{s}-\hat H_{k}^{(M)} |\chi_{k}\rangle_{s}=\frac{\exp\paq{{i\int^\eta \langle \tilde \chi_k|\hat H_{k}^{(M)}|\tilde \chi_k\rangle d\eta'}}}{2\M^2}\nonumber\\
\!\!\!\!\!\!\!&&\!\!\!\!\!\!\!\times\paq{\partial_a^2-\frac{a''}{(a')^{2}}\partial_{a}-\langle\tilde \chi_{k}|\partial_a^2-\frac{a''}{(a')^{2}}\partial_{a}|\tilde \chi_{k}\rangle} |\tilde\chi_k\rangle\equiv\epsilon \paq{\hat \Omega_{k}-\av{\hat \Omega_{k}}_{s}}|\chi_{k}\rangle_{s}
\ea
where $\av{\hat O}_{s}\equiv \phantom|_{s}\langle \chi_{k} |\hat O|\chi_{k}\rangle_{s}$ and $\epsilon\equiv\frac{1}{2\M^{2}}$. In Eq. (\ref{evoeq}) we have retained all terms in order to consistently include contributions to $\mathcal{O}\pa{\M^{-2}}$ (different expansions have been previously examined and compared for the homogeneous case \cite{BO-unit}). 
The operator $\hat \Omega_{k}$ has the following form:
\be{Omega}
\hat \Omega_{k}=\frac{1}{a'^{2}}\frac{d^{2}}{d\eta^{2}}+\paq{2i\frac{\av{\hat H_{k}^{(M)}}_{s}}{a'^{2}}-2\frac{a''}{a'^{3}}}\frac{d}{d\eta}.
\ee
The operator on the r.h.s. of Eq. (\ref{evoeq}) has a nonlinear structure since it depends on $\chi_{s}$ and $\chi_{s}^{*}$ through multiplicative factors of the form $\av{\hat O}_{s}$. 
We immediately note that in the absence of fluctuations (zero r.h.s.) Eq. (\ref{evoeq}) becomes the usual matter evolution equation (Schr\"odinger or Schwinger-Tomonaga). The terms on the r.h.s. describe the non-adiabatic effects of quantum gravitational origin.\\
For each $k$ mode, on neglecting such quantum gravity effects, Eq. (\ref{evoeq}) takes the form of a time dependent Schr\"odinger equation for a harmonic oscillator with time dependent frequency
\be{TDH}
\hat H_{k}^{(M)}=\frac{\hat \pi_{k}^{2}}{2}+\frac{\omega_{k}^{2}}{2}\hat v_{k}^{2}.
\ee
Its solutions can be generated by the linear invariant operator $\hat I$ which satisfies the equation
\be{inveq}
i \frac{d}{d\eta}\hat I+\paq{\hat I,\hat H}=0
\ee
where the subscript $k$ and the label $(M)$ have been removed and will henceforth be omitted.
The linear invariant can be written in terms of the Ermakov-Pinney variable \cite{Erm-Pin} $\rho$ as
\be{InvP}
\hat I=\frac{e^{i\Theta}}{\sqrt{2}}\paq{\pa{\frac{1}{\rho}-i\rho'}\hat v+i\rho\hat \pi}
\ee
with 
\be{pin}
\rho''+\omega^{2}\rho=\frac{1}{\rho^{3}}
\ee
and $\Theta=\int^{\eta} \frac{d\eta'}{\rho^{2}}$. The Bunch-Davies vacuum $|{\rm vac}\rangle$ is annihilated by $\hat I$ ($\hat I|{\rm vac}\rangle=0$) and excited states are created by $\hat I^\dagger$ acting on the vacuum \cite{BO-cosm}. In the coordinate representation the properly normalized vacuum is
\be{vacBD}
\langle v|{\rm vac}\rangle=\frac{1}{\pa{\pi\rho^2}^{1/4}}\exp\paq{\frac{i}{2}\int^\eta \frac{d\eta'}{\rho^2}-\frac{v^2}{2}\pa{\frac{1}{\rho^2}-i\frac{\rho'}{\rho}}}.
\ee
Instead, when quantum gravitational effects are taken into account, one must solve the integro-differential equation (\ref{evoeq}) which is an extremely difficult task.\\

We are interested in the spectrum of the scalar fluctuations which can be calculated from the two-point function 
\be{two}
p(\eta)\equiv \!\!\!\!\phantom{A}_{s}\langle 0|\hat v^{2}|0\rangle_{s}=\av{\hat v^{2}}_{0}
\ee
and can be compared with observations. The vacuum $|0\rangle_{s}$ satisfies the full equation (\ref{mateqk}) and should reduce to the BD vacuum in the short wavelength regime. Instead of trying to solve (\ref{evoeq}) and then calculating the power spectrum, one can find the differential equation for the spectrum $p$ by iteratively differentiating the two-point function and using the commutation relations.\\
On taking $|\chi_{k}\rangle_{s}=|0\rangle_{s}$ in Eq. (\ref{evoeq}) (we are omitting the subscript $k$) one obtains the evolution equation for the vacuum
\ba{evo0}
\!\!\!0&=&\!\!\!i\frac{d}{d\eta}|0\rangle_{s}-\hat H|0\rangle_{s}-\left[\left(2i\av{\hat H}_{0} g\pa{\eta}+g'\pa{\eta}\right)\right.\nonumber\\
\!\!\!&\times&\!\!\!\left.\pa{\frac{d}{d\eta}-\av{\frac{d}{d\eta}}_{0}}+g(\eta)\pa{\frac{d^{2}}{d\eta^{2}}-\av{\frac{d^{2}}{d\eta^{2}}}_{0}}\right]|0\rangle_{s}
\ea
with $\av{\hat O}_{0}\equiv \!\!\!\!\phantom{A}_{s}\langle 0|\hat O|0\rangle_{s}$ and $g(\eta)=\frac{1}{2\M^{2}a'^{2}}$.
The evolution of the two-point function can be now calculated by differentiating (\ref{two}) w.r.t. $\eta$ and using (\ref{evo0}). The first derivative of $p$ w.r.t. the conformal time is
\be{p1}
i\frac{dp}{d\eta}=\av{\left[\hat v^{2},\hat H\right]}_{0}-\av{\hat v^{2}}_{0}F(\eta)+G_{\hat v^{2}}(\eta)
\ee
where
\be{Fdef}
F(\eta)=\pa{2i g\av{\hat H}_{0}+g'}\av{\partial_{\eta}}_{0}+g\av{\partial_{\eta}^{2}}_{0}-c.c.\;,
\ee
\be{Gdef}
G_{\hat v^{2}}(\eta)=\pa{2i g\av{\hat H}_{0}+g'}\av{\hat v^{2}\partial_{\eta}}_{0}+g\av{\hat v^{2}\partial_{\eta}^{2}}_{0}-c.c..
\ee
Let us note that $g$ is a real function and $F$ and $G_{\hat v^{2}}$ are then purely imaginary functions of $\eta$ by construction. The subscript $\hat v^{2}$ in (\ref{Gdef}) indicates that the function $G$ depends on $\eta$ and on the operator $\hat v^{2}$. 
The commutator in the expression (\ref{p1}) is $\left[\hat v^{2},\hat H\right]=i\left\{\hat v,\hat \pi\right\}$.
In a more compact form Eq. (\ref{p1}) can then be written as
\be{p1com}
\frac{d\av{\hat v^{2}}_{0}}{d\eta}=\av{\left\{\hat v,\hat \pi\right\}}_{0}-iR(\hat v^{2})
\ee
where $R$ contains the quantum gravitational effects and is defined as $R(\hat O)=-\av{\hat O}_{0}F(\eta)+G_{\hat O}(\eta)$.
The above expression can be differentiated once more w.r.t. $\eta$ and takes the following form
\be{p2}
\frac{d^{2}\av{\hat v^{2}}_{0}}{d\eta^{2}}=\frac{d\av{\left\{\hat v,\hat \pi\right\}}_{0}}{d\eta}-i\frac{d R(\hat v^{2})}{d\eta}.
\ee
and, in analogy with (\ref{p1}) 
\be{p2term}
\frac{d\av{\left\{\hat v,\hat \pi\right\}}_{0}}{d\eta}=-i\av{\left[\left\{\hat v,\hat \pi\right\},\hat H\right]}_{0}-iR\left(\left\{\hat v,\hat \pi\right\}\right).
\ee
The commutator in the expression above becomes $\left[\left\{\hat v,\hat \pi\right\},\hat H\right]=2i\pa{\hat \pi^{2}-\omega^{2}\hat v^{2}}$ and (\ref{p2}) can be then rewritten as
\be{p2com}
\frac{d^{2}\av{\hat v^{2}}_{0}}{d\eta^{2}}=2\pa{\av{\hat \pi^{2}}_{0}-\omega^{2}\av{\hat v^{2}}_{0}}-iR\left(\left\{\hat v,\hat \pi\right\}\right)-i\frac{d R(\hat v^{2})}{d\eta}.
\ee
On then calculating  the derivative of  Eq. (\ref{p2com}) we finally obtain:
\ba{p3}
\frac{d^{3}\av{\hat v^{2}}_{0}}{d\eta^{3}}\!\!\!&=&\!\!\!\frac{d\av{\hat \pi^{2}}_{0}}{d\eta}-4\omega\omega'\av{\hat v^{2}}_{0}-2\omega^{2}\frac{d\av{\hat v^{2}}_{0}}{d\eta}\nonumber\\
\!\!\!&-&\!\!\!i\frac{dR\left(\left\{\hat v,\hat \pi\right\}\right)}{d\eta}-i\frac{d^{2} R(\hat v^{2})}{d\eta^{2}},
\ea	
where
\be{com3}
\frac{d\av{\hat \pi^{2}}_{0}}{d\eta}+iR(\hat \pi^{2})=-i\av{\left[\hat \pi^{2},\hat H\right]}_{0}
=i\omega^{2}\av{\left[\hat v^{2},\hat H\right]}_{0}
\ee
and 
\be{com4}
\av{\left[\hat v^{2},\hat H\right]}_{0}=i\frac{d\av{\hat v^{2}}_{0}}{d\eta}-R(\hat v^{2}).
\ee
Equation (\ref{p3}) finally becomes
\ba{p33}
0\!\!\!&=&\!\!\!\frac{d^{3}\av{\hat v^{2}}_{0}}{d\eta^{3}}+4\omega^{2}\frac{d\av{\hat v^{2}}_{0}}{d\eta}+2\pa{\omega^{2}}'\av{\hat v^{2}}_{0}+2iR(\hat \pi^{2})\nonumber\\
\!\!\!&+&\!\!\!2i\omega^{2}R(\hat v^{2})+i\frac{dR\left(\left\{\hat v,\hat \pi\right\}\right)}{d\eta}+i\frac{d^{2} R(\hat v^{2})}{d\eta^{2}}.
\ea

The solution of the homogeneous equation is $\rho^{2}/2$ where $\rho$ is the Ermakov-Pinney variable \cite{Erm-Pin}. The remaining terms describe the quantum gravitational effects on the evolution of the power spectrum. Let us note that for a Hermitian operator $\hat O$, $R(\hat O)$
is a purely imaginary function of $\eta$. Therefore the quantum gravitational contributions in (\ref{p33}) are real.\\
Again solving the integro-differential equation (\ref{p33}) exactly is complicated but its very structure is suitable for a perturbative approach, at least to first order in $1/\M^{2}$. Given the precision of the present status of cosmological observations, the first order solution is sufficient in order to seek quantum gravity effects by comparison with the data. The perturbed solution of (\ref{p33}) can be obtained by estimating all the terms which contain $R(\hat O)$ perturbatively by using the vacuum state of the unperturbed evolution (\ref{vacBD}). This is sufficient to obtain the quantum gravitational corrections to order $1/\M^{2}$.\\
For a general background the effects of quantum gravity can then be written in terms of $\rho^{2}$, namely the solution of the homogeneous equation, or equivalently in terms of $p$ itself which coincides with $\rho^{2}/2$ to order $1/\M^{2}$.
Then Eq. (\ref{p33}) has the following form:
\ba{qfx}
\!\!\!&&\!\!\!\frac{d^{3}p}{d\eta^{3}}+4\omega^{2}\frac{dp}{d\eta}+2\frac{d\omega^{2}}{d\eta}p-\frac{1}{\M^{2}}\frac{d^{3}}{d\eta^{3}}\frac{\left(p'^{2}+4\omega^{2}p^{2}-1\right)}{4 a'^{2}}\nonumber\\
\!\!\!&&\!\!\!+\frac{1}{\M^{2}}\frac{d^{2}}{d\eta^{2}}\frac{p'\left(p'^{2}+4\omega^{2}p^{2}+1\right)}{4 pa'^{2}}
+\frac{1}{\M^{2}}\frac{d}{d\eta}\left\{\frac{1}{8a'^{2}p^{2}}\left[\left(1-4\omega^{2}p^{2}\right)^{2}\right.\right.\nonumber\\
\phantom{\frac{A}{B}}\!\!\!&&\!\!\!\left.\left.\left.+2p'^{2}\left(1+4\omega^{2}p^{2}\right.\right)+p'^{4}\right]\right\}-\frac{1}{\M^{2}}\frac{\omega\omega'\left(p'^{2}+4\omega^{2}p^{2}-1\right)}{a'^{2}}=0
\ea
The explicit form of this equation represents the main result of this letter. Such an equation may be used in the following way: i) one first chooses a classical background evolution obtaining a particular form for $\omega$ and $g$; ii) one then solves the
unperturbed Eq. (\ref{qfx}) (i.e. the equation without the terms proportional to $1/M_P^2$), choosing  the integration constants so as to have $p = 1/2k$
in the short wavelength limit. One then substitutes the chosen solution of the homogeneous equation into the terms proportional to
$1/M_P^2$ and finally solves the resulting inhomogeneous equation for $p$.

\section{Application to de Sitter and Conclusions}
In viable single field inflationary models one has an evolution of cosmological perturbations based on the slow-roll paradigm. One then has a quasi de Sitter expansion, small slow-roll parameters, a nearly flat spectrum and a finite amplitude of scalar fluctuations. For such a case, in order to illustrate the main effects of quantum gravity on the spectrum it will be sufficient to neglect slow-roll parameters, that is just consider a pure de Sitter expansion. \\
When $H = \rm{const}$, one has $\omega=\sqrt{k^{2}-\frac{2}{\eta^{2}}}$. 
The solution of Eq. (\ref{pin}) compatible with the short wavelength flat spacetime limit is $\rho_{DS}=\sqrt{\frac{1}{k}+\frac{1}{k^{3}\eta^{2}}}$ then Eq. (\ref{qfx}) takes the very simple form:
\be{pinpert}
\frac{d^{3}p}{d\eta^{3}}+4\pa{k^{2}-\frac{2}{\eta^{2}}}\frac{dp}{d\eta}+\frac{8}{\eta^{3}}p+\frac{4H^{2}}{\M^{2}k^{4}\eta^{3}}=0.
\ee
The full solution for $p$ is
\ba{fullsol}
p&=&\frac{1}{2k^{4}\eta^{2}}\left\{c_{+}\pa{1+k^{2}\eta^{2}}+\cos\pa{2k\eta}\paq{2c_{0}k\eta-c_{-}\pa{k^{2}\eta^{2}-1}}\right.\nonumber\\
&&\left.+\sin\pa{2k\eta}\left[c_{0}\pa{k^{2}\eta^{2}-1}+2c_{-}k\eta
\right]-\frac{H^{2}}{\M^{2}}\eta^{2}\right\}
\ea
To reproduce the solution $\rho_{DS}$ in the absence of the quantum gravitational effects one must choose $c_{-}=c_{0}=0$ and $c_{+}=k$ to obtain:
\be{spectrumfull}
\mathcal{P}_{v}=\frac{k^{3}}{2\pi^{2}}p=\frac{a^2H^2}{4\pi^2}\pa{1+\frac{k^2}{a^2H^2}-\frac{1}{a^2k\M^2}}.
\ee  
In the long wavelength limit ($-k\eta\rightarrow 0$) the solution (\ref{spectrumfull}) leads to the final result for the spectrum 
\be{spectrum}
\mathcal{P}_{v}\stackrel{-k\eta\rightarrow 0}{=}\frac{a^{2}H^{2}}{4\pi^{2}}\pa{1-\frac{1}{a^{2}\M^{2}k}}.
\ee
We observe that, with such a choice of initial conditions, the quantum gravitational correction leads
to a running of the spectral index and less power for large scales. This would be in qualitative agreement with Planck results \cite{cmb}.
However the quantum gravitational correction in (\ref{spectrumfull}) must be small for the perturbative approach to be valid. Further it should dominate over the term $\frac{k^2}{a^2H^2}$ in order for (\ref{spectrum}) to be valid (long wavelength limit). The latter requirements set the following constraints on such a correction and consequently on $k$:
\be{limcorr}
\frac{k^2}{a^2H^2}\ll \frac{1}{a^{2}\M^{2}k}\ll 1\Rightarrow \frac{1}{a^2\M^2}\ll k\ll \pa{\frac{H^2}{\M^2}}^{1/3}
\ee
Let us now note that the time dependence in Eq. (\ref{spectrumfull}) occurs only in the scale factor $a$ and such a time dependence is exactly the same in two of the 3 terms in the bracket. The second term describes the short wavelength limit of the spectrum in the absence of fluctuations, associated with the Bunch-Davies (BD) vacuum, and the third term corresponds to the quantum gravitational correction obtained from the non adiabatic effects. At some time during the primordial expansion such terms dominate over the constant term. For modes with $k\gg \pa{\frac{H^2}{\M^2}}^{1/3}$ the BD contribution becomes the leading one for $a(\eta)$ small enough but for the modes with $k\ll \pa{\frac{H^2}{\M^2}}^{1/3}$ (which correspond to the quantum gravitational deviations from the pure de Sitter power spectrum) the BD term is negligible w.r.t. the quantum gravitational term for $a(\eta)$ small. This is in contradiction of our perturbative evaluation of the quantum gravitational correction wherein we have assumed that a BD vacuum is the dominant contribution even in the presence of large quantum gravitational effects for very early times. It may well be that such a BD requirement is only valid for not so early times in which case our expression (\ref{spectrum}) may be valid.

On the other hand one can impose the the BD limit be respected for all modes $k$. This can be done by choosing the integration constants in (\ref{fullsol}) to be $c_{-}=c_{0}=0$ and $c_{+}=k+\pa{H^{2}/\M^{2}}c(k)$ which is compatible with Eq. (\ref{pinpert}) up to correction of the order $H^{4}/\M^{4}$ (which we have always omitted). We then obtain: 
\be{spfullmod}
\tilde\mathcal{P}_{v}=\frac{a^2H^2}{4\pi^2}\pa{1+\frac{k^2}{a^2H^2}-\frac{1}{a^2k\M^2}+\frac{c(k)}{k}\frac{H^2}{\M^2}+\frac{c(k)k^2}{a^2k \M^2}}.
\ee 
On setting $c(k)=1/k^2$ one can eliminate the quantum gravitational term which has the same time dependence as the BD vacuum contribution and leads to the difficulties  described above for small $k$ and $a(\eta)\rightarrow 0$. Now Eq. (\ref{spfullmod}) becomes 
\be{spfullmod2}
\tilde \mathcal{P}_{v}=\frac{a^2H^2}{4\pi^2}\pa{1+\frac{k^2}{a^2H^2}+\frac{1}{k^3}\frac{H^2}{\M^2}}\stackrel{-k\eta\rightarrow 0}{=}\frac{a^2H^2}{4\pi^2}\pa{1+\frac{1}{k^3}\frac{H^2}{\M^2}}.
\ee 
The spectrum (\ref{spfullmod2}) is qualitatively different from that in Eq. (\ref{spectrum}). The quantum gravitational correction now scales as $k^{-3}$ and leads to an increase of power for large scales. This may be a general result and has been found in other approaches (see \cite{modes},\cite{Kiefer},\cite{Esp-Kiefer}).

To summarize, we obtained a general equation, describing the lowest order corrections to the spectrum of the cosmological scalar fluctuations 
coming from quantum gravitational effects and then estimated these corrections explicitly for the case of a de Sitter evolution.
The influence of the quantum gravitational effects on the spectrum of fluctuations was considered recently in papers \cite{Kiefer,Esp-Kiefer}. The difference between the approach used in these  papers and our approach is actually discussed in the Appendix of paper \cite{Esp-Kiefer}. The crucial point is that we follow a decomposition of the total wave function in a purely gravitational and a matter part as is done in traditional BO approaches and retain the r.h.s. of the resulting equations. This corresponds to including the fluctuations due to non-adiabaticity associated with quantum gravitational effects. Further we note that our use of the gauge-invariant variables is necessary to study in a self-consistent way both the scalar perturbations of the metric and the scalar field fluctuations. In our approach, as previously pointed out \cite{BO-unit}, the unitarity problem never arises. 
The detailed explanation of our method and its application to other backgrounds will be presented elsewhere \cite{WIP}. \\

\section*{Acknowledgments}
A.K. was partially supported by the RFBR grant 11-02-00643. We wish to thank G. Esposito, C. Kiefer, M. Peloso, F. Pessina and G.P. Vacca for useful comments and suggestions.


\end{document}